\documentclass[prl,twocolumn]{revtex4}

\usepackage{graphicx}
\usepackage{dcolumn}
\usepackage{bm}

\begin{document}

\title{Optical Hall Effect in the Integer Quantum Hall Regime}

\author{Y. Ikebe}
\affiliation{Department of Physics, The University of Tokyo, Hongo, Tokyo 113-0033, Japan}
\author{T. Morimoto}
\affiliation{Department of Physics, The University of Tokyo, Hongo, Tokyo 113-0033, Japan}
\author{R. Masutomi}
\affiliation{Department of Physics, The University of Tokyo, Hongo, Tokyo 113-0033, Japan}
\author{T. Okamoto}
\affiliation{Department of Physics, The University of Tokyo, Hongo, Tokyo 113-0033, Japan}
\author{H. Aoki}
\affiliation{Department of Physics, The University of Tokyo, Hongo, Tokyo 113-0033, Japan}
\author{R. Shimano}
\affiliation{Department of Physics, The University of Tokyo, Hongo, Tokyo 113-0033, Japan}

\date{\today}

\begin{abstract}
Optical Hall conductivity $\sigma_{xy}(\omega)$ is measured from the Faraday rotation for a GaAs/AlGaAs heterojunction quantum Hall system in the terahertz frequency regime. The Faraday rotation angle ($\sim$ fine structure constant $\sim$ mrad) is found to significantly deviate from the Drude-like behavior to exhibit a plateau-like structure around the Landau-level filling $\nu=2$. The result, which fits with the behavior expected from the carrier localization effect in the ac regime, indicates that the plateau structure, although not quantized, still exists in the terahertz regime.
\end{abstract}

\maketitle

The quantum Hall effect (QHE), a highlight in the two-dimensional electron gas (2DEG) system in strong magnetic fields \cite{Prange,Laughlin,Aoki,Halperin}, still harbors, despite its long history, a wealth of important physics. While static properties of the integer QHE have been well understood, we are still some way from a full understanding of dynamical responses in the QHE in the ac or even optical regime. In the static case, the states localized due to disorder with the localization length smaller than the sample size or the inelastic scattering length are crucial in realizing the quantum plateaus for the dc Hall current in a dc electric field \cite{Aoki2,Pruisken,Huckestein,Chalker,Wei,Koch,Engel}. On the other hand, the conventional wisdom for the dynamical response would be that an ac field will delocalize wave functions to make QHE disappear.

For relatively low frequencies, the breakdown of QHE in ac fields has a long history of investigation \cite{Pepper}. One issue was whether the delocalization occurs for low-frequencies ($\sim 10\;\mathrm{MHz}$), but the results were not conclusive.  Subsequently, experimental study was extended to the microwave regime in the 1980s, where the delocalization as seen in the Hall conductivity $\sigma_{xy}$ was shown to be absent in the microwave (i.e., gigahertz) regime \cite{Kuchar}, while the gigahertz responses of the longitudinal conductivity $\sigma_{xx}$ \cite{Engel,Balaban,Hohls} were explained with the scaling theory of localization \cite{Abrahams}. Thus, a fundamental problem remains as to whether and how QHE is affected in the much higher, {\it terahertz} (closer to the optical) frequency regime ($\omega \sim 10^{12}\;\mathrm{Hz} \sim 10^{-2}\;\mathrm{eV}/\hbar$). This is an essential question, since the frequency is exactly the energy scale of interest (i.e., the cyclotron energy $\hbar\omega_c \sim 10^{-2}\;\mathrm{eV}$ for a magnetic field $\sim 10\;\mathrm{T}$, which is the spacing between Landau levels, a prerequisite for QHE).

\begin{figure}
\begin{center}
\includegraphics[width=85mm]{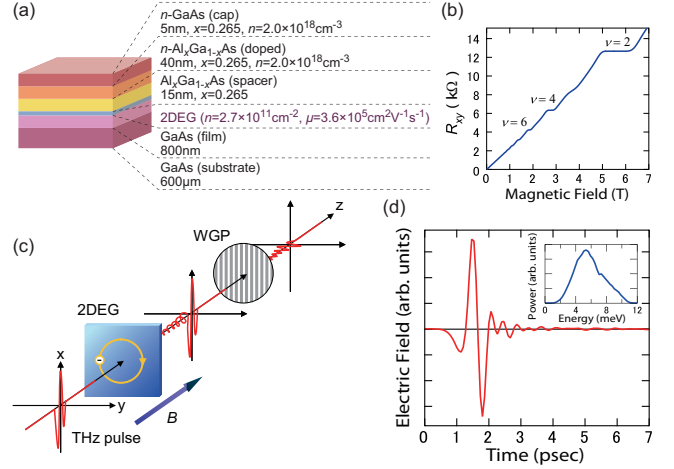}
\end{center}
\caption{(color online) (a) Schematic sample geometry. A modulation doped GaAs/AlGaAs single heterojunction is used as a 2DEG system.
(b) Magnetic-field dependence of the dc Hall resistivity, $R_{xy}$.
Plateau structures in the QHE numbers of 2, 4, and 6 are indicated around  $5.6\;\mathrm{T}$, $2.8\;\mathrm{T}$ and $1.9\;\mathrm{T}$, respectively.
(c) Schematic configuration for the Faraday rotation measurements in a Cross-Nicole geometry. 
The polarization state of the terahertz wave changes from linear to elliptical after transmitting the sample due to the cyclotron motion of the 2DEG electrons. WGP: wire grid polarizer.
(d) Time-domain waveform of the electric field of the incident terahertz pulse. Inset: power spectrum of the terahertz pulse, ranging from $1.2$ to $10\;\mathrm{meV}$ ($0.3$ to $2.5\;\mathrm{THz}$).}
\end{figure}

Theoretically, the accurate quantization in QHE is firmly established as a topological (Chern) number \cite{Thouless} in the static case. However, such a picture may not be extended to the ac regime where the topological `protection' no longer exists. Recently, Morimoto $et\;al.$ \cite{Morimoto} have theoretically examined the ac response of the disordered QHE systems based on the exact diagonalization method, and showed that a plateau-like behavior still exists in $\sigma_{xy}$ even in the terahertz energy range. This has motivated us to experimentally examine QHE by going beyond the microwave regime, which has so far remained a challenge. An essential experimental ingredient that enables the measurement is a recent development in terahertz time-domain spectroscopy (THz-TDS) \cite{Nuss,Mittleman,Hangyo}. By combining the polarization spectroscopy with THz-TDS, the magneto-optical Faraday effect and Kerr effect have become measurable in high-$T_c$ superconductors \cite{Parks} and in doped semiconductors \cite{Mittleman2,Shimano}. Since these magneto-optical effects are high frequency counterparts to the dc Hall effect, we are now plunging into the ``optical Hall conductivity" as a novel probe.
In the present Letter, we have investigated the terahertz-frequency Hall conductivity in 2DEG in the quantum Hall regime to report an evidence for a quantum Hall plateau in $\sigma_{xy}$ in the terahertz frequency regime. In the plateau region of $\sigma_{xy}(\omega)$, the Faraday rotation angle, detected in a transmission configuration, is of the order of the fine-structure constant ($\sim$ mrad).

Our sample consisted of a modulation-doped GaAs/AlGaAs single heterojunction (Fig.~1~(a)), whose lateral size was $1\times 1\;\mathrm{cm}^2$, large enough to cover the spot size ($1.2\;\mathrm{mm}$ in diameter at $1\;\mathrm{THz}$) of the terahertz probe. The dc Hall resistivity $R_{xy}$ at $T=3\;\mathrm{K}$ (Fig.~1~(b)) shows a plateau structure around the magnetic field of $5.6\;\mathrm{T}$ corresponding to the QHE number of 2. From the slope of $R_{xy}$ in low magnetic fields, the carrier density in the 2DEG layer was determined as $n=2.7\times 10^{11}\;\mathrm{cm}^{-2}$.

\begin{figure}
\begin{center}
\includegraphics[width=85mm]{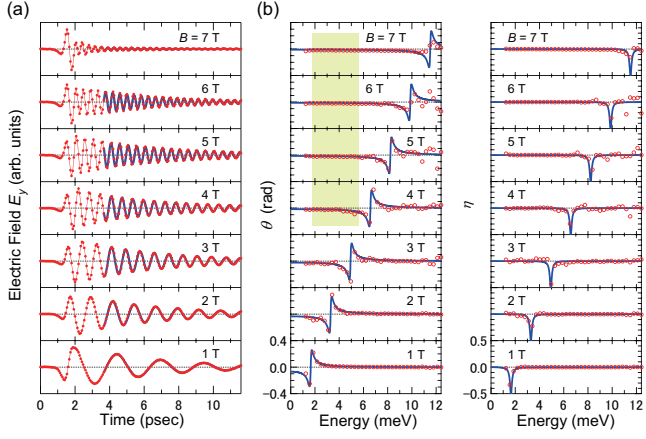}
\end{center}
\caption{(color online) (a) Time-domain waveforms of the transmitted terahertz electric field in $y$-polarization measured at $T=3\;\mathrm{K}$ at the indicated magnetic fields. Fits to the equation $E_y\left(t\right)\propto\exp(-t/\tau)\sin\left(\omega_c t\right)$, as shown by bold lines, give the mobility of $\mu=1.4\times10^5\;\mathrm{cm^2/Vs}$ and the effective mass of $m^*=0.070m_0$ ($m_0$ is the electron mass), respectively.
(b) The corresponding Faraday rotation angle $\theta$ and the ellipticity $\eta$ (open circles) obtained from the Fourier transformation of Fig.~2~(a) (see text). Solid lines: $\theta$ and $\eta$ calculated from Drude-like classical limit, eqs.~(2,3). The hatched region in Fig.~2~(b) indicates the region considered in Fig.~3.}
\end{figure}

The configuration for the Faraday rotation measurements is schematically shown in Fig.~1~(c). A linearly ($x$-)polarized and nearly single-cycle terahertz pulse (Fig.~1~(d)) is irradiated onto the sample in a magnetic field. The power spectrum shown in the inset of Fig.~1~(d) ensures a measurement frequency range from $1.2$ to $10\;\mathrm{meV}$ ($0.3$ to $2.5\;\mathrm{THz}$). Because of the Faraday effect in the 2DEG layer, an orthogonal ($y$-)component of the electric field is induced in the transmitted pulse. We then measured the transmitted terahertz $E_x(t)$ and $E_y(t)$ respectively, by rotating the wire grid polarizer (WGP) placed beyond the sample. We have then obtained the Faraday rotation angle $\theta$ and the ellipticity $\eta$ from the Fourier transformation of $E_{x(y)}(t)$ by using the relation
\begin{equation}
\frac{E_y(\omega)}{E_x(\omega)}=\frac{\sin\theta+i\eta\cos\theta}{\cos\theta-i\eta\sin\theta}.
\end{equation}
With this method we were able to detect $\theta$ as small as $0.5\;\mathrm{mrad}$. Further experimental details are described in Ref. \cite{Ikebe}.

Figure~2~(a) displays the magnetic-field dependence of $E_y(t)$ at $T=3\;\mathrm{K}$. Clear oscillations that are sustained much longer than the incident terahertz pulse width are observed. The oscillation frequency monotonically increases with the magnetic field as expected for the cyclotron frequency $\omega_c=eB/m^*$ ($m^*$: the effective mass). The damped oscillatory behavior corresponds to the free induction decay (FID) of the cyclotron motion of electrons as expressed by $E_y\left(t\right)\propto \exp(-t/\tau)\sin\left(\omega_c t\right)$, where the scattering time $\tau$ is related to the electron mobility $\mu$ as $\tau=m^*\mu/e$ \cite{footnote}. The long-lived oscillatory signal indicates a high mobility of the 2DEG system. With $\tau$ and $\omega_c$ obtained from the fitting, we have evaluated $\mu=1.4\times10^5\;\mathrm{cm^2/Vs}$ and $m^*=0.070m_0$ ($m_0$: the bare electron mass).

The complex Faraday rotation spectrum $\Theta(\omega)=\theta(\omega)+i\eta(\omega)$ is then obtained with eq.~(1).  The result in Fig.~2~(b) shows that $\theta$ and $\eta$ respectively exhibit sharp structures around the cyclotron resonance frequency. The width of the cyclotron resonance is much smaller than our frequency resolution of $0.075\;\mathrm{THz}$, reflecting the high electron mobility. To elucidate the Hall conductivity from the observed Faraday rotation spectra, we have adopted a thin-film limit approximation \cite{OConnell}, 
\begin{equation}
\theta(\omega)+i\eta(\omega) \simeq \frac{1}{(1+n_{\mathrm{sub}})c\epsilon_0}\sigma_{xy}(\omega),
\end{equation}
where $n_{\mathrm{sub}}$ is the refractive index of the substrate, $c$ the speed of  light and $\epsilon_0$ the vacuum permittivity.  As far as the resonance lineshape is concerned, $\sigma_{xy}(\omega)$ is expressed, in the Drude model, as
\begin{equation}
\sigma_{xy}(\omega) \simeq \frac{ne^2}{m^*}\frac{\omega_c}{(\omega+i/\tau)^2-\omega_c^2}.
\end{equation}
The solid lines in Fig.~2~(b) represent the curves calculated from eqs.~(2,3), which agree well with the experimental results. 

In the optical Hall conductivity, a numerical result \cite{Morimoto} indicates that the overall dependence of $\sigma_{xy}$ as a function of two variables, $\omega$ and the carrier concentration $n$, is such that the $n$-dependence retains a plateau-like structure even for each finite value of $\omega$, while the $\omega$-dependence takes a cyclotron-resonance-like line shape, which becomes very accurate in the tail regions of the cyclotron resonance. So here we define a normalized optical Hall conductivity to remove the $\omega$-dependence coming from the cyclotron resonance as
\begin{equation}
\tilde{\sigma}_{xy}(\omega) \equiv  \left[\frac{e^2}{h}\frac{\omega_c^2}{(\omega+i/\tau)^2-\omega_c^2}\right] ^{-1} \sigma_{xy}(\omega).
\end{equation}
Since $\tilde\sigma_{xy}(\omega)$ plays the role of an effective $n$ replacing $n$ in eq.~(3), we can look for a plateau in $\tilde{\sigma}_{xy}(\omega)$ around an even integer of the Landau level filling $\nu$ where spin-degenerated Landau level is filled. In terms of the Faraday rotation this reads, via eqs.~(2,4), $\theta(\omega)$ approaching to $-2N\alpha/(1+n_{\mathrm{sub}}) (\simeq \mathrm{mrad})$ in the low-frequency region of $\omega\ll\omega_c$, where $\alpha=e^2/(4\pi\epsilon_0\hbar c)$ is the fine-structure constant and $N$ is the quantized integer. We display both $\theta(\omega)$ (Fig.~3~(a)) and $\tilde{\sigma}_{xy}(\omega)$ (Fig.~3~(b)) against energy from $2\;\mathrm{meV}$ to $5.3\;\mathrm{meV}$, which corresponds to the frequency range below the cyclotron resonance at $\omega_c$ for the magnetic field range around $\nu=2$ ($B=5.6\;\mathrm{T}$) as indicated in Fig.~2~(b) \cite{footnote2}. Also shown are the theoretical curves in the Drude-like classical limit, eq.~(3), along with the quantum Hall limit, $\tilde{\sigma}_{xy}=2$. For these curves, we have employed the experimentally determined parameters: the carrier density obtained from the dc Hall measurement, and $\tau$ obtained from the FID \cite{footnote3}. We can see in Fig.~3~(b) that the experimental $\tilde{\sigma}_{xy}(\omega)$  exhibits reasonably flat behavior, which indicates that the normalization we have adopted satisfactory captures the $\omega$-dependence.

\begin{figure}
\begin{center}
\includegraphics[width=85mm]{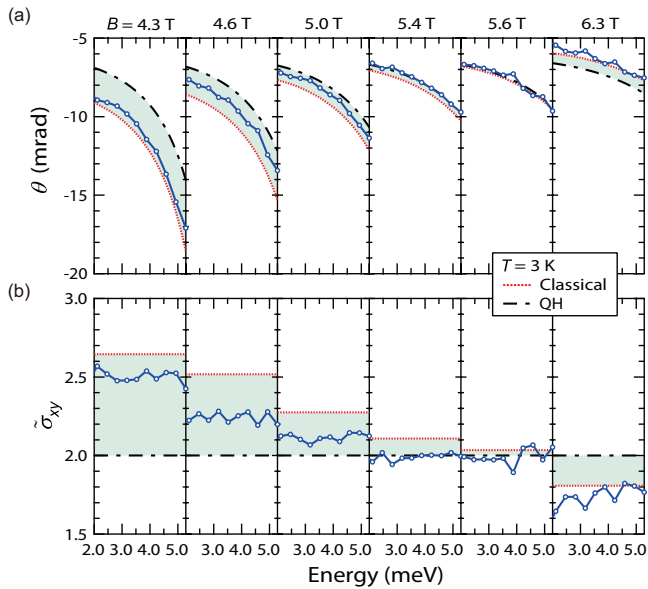}
\end{center}
\caption{(color online) (a) Experimental spectra (dots) for the Faraday rotation angle, $\theta(\omega)$, at $T=3\;\mathrm{K}$ and at the indicated magnetic fields around $B=5.6\;\mathrm{T}$ ($\nu=2$).
(b) Similar plot for the normalized Hall conductivity $\tilde{\sigma}_{xy}(\omega)$. Theoretical curves calculated from eq.~(4) with the Drude-like classical limit (eq.~(3); dotted lines) and from the quantum Hall limit (with $nh/(eB)$ taken to be quantized integers $N$; dashed lines) are also shown with their difference highlighted by hatching.}
\end{figure}

 This enables us to focus on the dependence on the carrier concentration, or the Landau level filling $\nu$.  We can see that, as $\nu$ is varied for the magnetic field increased from $4.3\;\mathrm{T}$ to $5.4\;\mathrm{T}$, the observed Faraday rotation angle significantly deviates from the Drude behavior, and approaches to the quantum one. Specifically, the observed Faraday rotation angle in the low-frequency edge of $2\;\mathrm{meV}$ takes a nearly constant value of $-6.3\;\mathrm{mrad}$ in the magnetic-field range between $5.2\;\mathrm{T}$ and $5.6\;\mathrm{T}$, in accordance with the value of $-4\alpha/(1+n_{\mathrm{sub}})=-6.3\;\mathrm{mrad}$ with $n_{\mathrm{sub}}=3.6$ for the GaAs substrate \cite{Johnson}.

So let us now plot the Faraday rotation angle represented as $\tilde{\sigma}_{xy}(\omega)$ against the Landau level filling $\nu \propto 1/B$ in Fig.~4~(a) in the frequency range indicated in Fig.~3 at two values of temperature, $T=3\;\mathrm{K}$ and $T=20\;\mathrm{K}$. The dotted line again represents the classical limit: $\tilde{\sigma}_{xy}(\omega) \rightarrow nh/(eB)$ with the carrier density $n=2.7\times 10^{11}\;\mathrm{cm}^{-2}$ again, while the quantum Hall limit by the dashed line. The result obtained from the dc Hall resistivity is also shown by the solid red line. Clearly, $\tilde{\sigma}_{xy}(\omega)$ at $T=3\;\mathrm{K}$ shows a marked deviation from the classical one for magnetic field $4.3\;\mathrm{T}< B < 5.4\;\mathrm{T}$, and exhibits a plateau-like behavior, precisely around $\nu=2$, which is more clearly seen in an enlarged plot in Fig.~4~(b). At a higher $T=20\;\mathrm{K}$, on the other hand, the plateau structure disappears and a good agreement with the classical limit is recovered.

The plateau hallmarks a QHE-like structure that remains even in the terahertz frequency range, but the plateau region is narrower and shifted to the lower magnetic field side of the dc plateau center at $B=5.6\;\mathrm{T}$ ($\nu=2$). Since the Fermi energy moves from the second Landau index to the lowest one at $5.6\;\mathrm{T}$, the observed asymmetric behavior may be attributed to the difference in the Landau-level broadening and in the localization length $\xi$ between the lowest and second Landau indices. According to the scaling theory, the localization length $\xi(E)$ diverges around the center of each Landau level, $E_c$, as $\xi(E)\propto|E-E_c|^{-s}$ ($s$: a critical exponent)  \cite{Aoki2,Pruisken,Huckestein,Chalker}, where a lower Landau level has a smaller $s$ \cite{Aoki2}. For dc Hall conductivity, the states with $\xi$ larger than the sample size $L$ contribute to the transport while the localized states with $\xi<L$ give rise to the plateau structure. Hence the extended states only reside in the very vicinity of the center of the Landau level since the cutoff is the large sample size $L$, and $\sigma_{xy}$ is not so sensitive to the difference in $s$.  In ac Hall conduction, by contrast, we have another length scale $\tilde L(\omega)$, the distance over which an electron travels in one cycle of the ac field, and the extended states with $\xi>\tilde L(\omega)$ contribute to $\sigma_{xy}(\omega)$. Since $\tilde L(\omega)$ is much smaller than $L$ in general, the plateau width is determined in the tail region of the localization length divergence, which tends to make the plateau shrink faster for Landau levels with a smaller $s$. This implies that the ac conductivity $\sigma_{xy}(\omega)$ should more sensitively reflect the difference in $s$ between the adjacent Landau levels. Indeed, when we numerically calculate $\sigma_{xy}(\omega)$ as a function of the carrier density with the exact diagonalization method \cite{Morimoto} which takes account of this difference in $s$, the result in Fig.~4~(c) exhibits an asymmetric behavior in the plateau structure.

\begin{figure}
\begin{center}
\includegraphics[width=85mm]{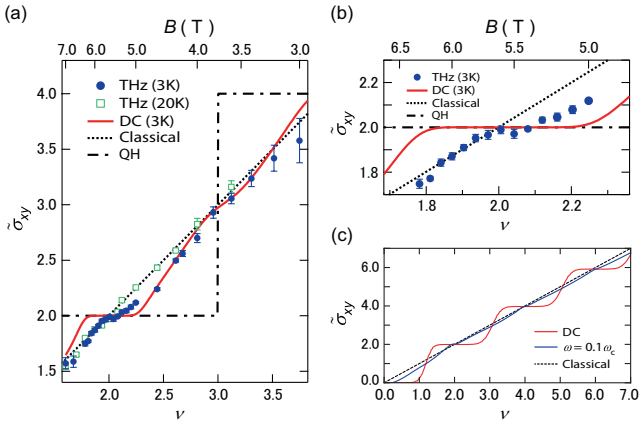}
\end{center}
\caption{(color online) (a) Magnetic-field dependence of the normalized Hall conductivity $\tilde{\sigma}_{xy}(\omega)$ determined from the fitting of the experimentally observed Faraday rotation angle at $T=3\;\mathrm{K}$ (solid circles) and $20\;\mathrm{K}$ (open squares). The theoretical values calculated from the Drude-like classical limit with $\tilde{\sigma}_{xy}=nh/(eB)$ and from the quantum Hall limit, $\tilde{\sigma}_{xy}=2$ and 4, are also shown by dotted and dashed lines, respectively. Solid curve represents the experimental dc Hall conductivity $\sigma_{xy}$.
(b) Enlarged Fig.~4~(a) around $\nu=2$.
(c) Theoretical result for $\tilde{\sigma}_{xy}(\omega)$ at $\omega=0$ (red line) and at a typical $\omega=0.1\omega_c$ (blue) plotted here as a function of the Landau level filling $\nu$ with the exact diagonalization method calculated here for a system size $L=40\ell$ with $\ell$ being the magnetic length and a Landau level broadening $\Gamma=\hbar\omega_c$ with $\tau=0$ for clarity. To facilitate comparison, we have included the spin degeneracy of 2 in $\nu$ and $\tilde{\sigma}_{xy}$.}
\end{figure}

In conclusion, we have investigated the Hall conductivity in a 2DEG in a heterojunction GaAs/AlGaAs in the terahertz frequency range by using the Faraday effect. Long-lived FID signal is observed in the transmitted terahertz wave, from which the Faraday rotation angle $\theta(\omega)$ is obtained. Around the filling of $\nu=2$, $\theta(\omega)$ takes a nearly constant value of $-4\alpha/(1+n_{\mathrm{sub}})=-6.3\;\mathrm{mrad}$ and a plateau behavior in the Hall conductivity is clearly observed even in the terahertz frequency range. We note that the terahertz physics as described here not only brings a new insight into the carrier dynamics in 2DEG in the quantum Hall regime, but will potentially offer a novel probe for studying the transport properties of materials governed by a topological nature of the electron wavefunction.

This work was partially supported by grants-in-aide (No. 09J09833, 20340098), Global COE Program ``the Physical Sciences Frontier", from MEXT, Japan, and PRESTO, Japan Science and Technology Agency.


\begin{thebibliography}{99}
\bibitem{footnote}
The damped oscillatory behavior is also seen for $E_x(t)$ with its phase shifted from that of $E_y(t)$, and both of $E_x(t)$ and $E_y(t)$ bare plugged into eq.~(1) to obtain the Faraday rotation spectra.
\bibitem{footnote2}
Here we have repeated the time-scan in THz-TDS to improve the signal to noise ratio and the accuracy of the rotation angle as compared to Fig.~2~(b).
\bibitem{footnote3}
From dc Hall measurements, we have estimated the drift mobility as $\mu=3.6\times10^5\;\mathrm{cm^2/Vs}$, which is larger than the mobility at THz region. While the detailed analysis of frequency dependence of the mobility will be reported elsewhere, the calculated $\theta$ is not sensitive to the value of the mobility for the frequency range $\omega_c-\omega\gg1/\tau$ (Fig.~3).
\end{thebibliography}
\end{document}